\documentclass{article}

\usepackage{txfonts}

\usepackage[utf8]{inputenc}
\usepackage[T1]{fontenc}
\usepackage[dvipsnames]{xcolor}

\usepackage{graphicx}	
\usepackage{float}

\usepackage{xspace}

\usepackage[normalem]{ulem}
\usepackage{siunitx}

\usepackage{fontawesome}

\usepackage{color}
\usepackage{tikz}
\usepackage{pgf-umlsd}
\usepackage{ifthen}
\usepackage{url}

\usepackage{hyperref}
\hypersetup{colorlinks=true, linkcolor=black, citecolor = blue, urlcolor=blue}

\usepackage{cleveref}
\newcommand{\github}{\href{https://github.com/BastienArcelin/IPU-GPU}{\faGithub}}
\newcommand{\githubimggen}{\href{https://github.com/BastienArcelin/image_generation_GalSim}{\faGithub}}
\crefformat{footnote}{#2\footnotemark[#1]#3}

\newcommand*{\ie}{i.e.\@\xspace}
\usepackage[backend=biber, natbib=true, style=authoryear,bibstyle=authortitle]{biblatex}
\usepackage{authblk}
\title{Comparison of Graphcore IPUs and Nvidia GPUs for cosmology applications}
\author[Bastien Arcelin]{Bastien Arcelin$^{1}$\thanks{E-mail: arcelin@apc.in2p3.fr (APC)}
\\
$^{1}$\textit{Université de Paris, CNRS, Astroparticule et Cosmologie, F-75013 Paris, France
}}
\date{}

\begin{document}
\label{firstpage}
\maketitle
\bigbreak

\noindent
\textbf{Abstract}. This paper represents the first investigation of the suitability and performance of Graphcore Intelligence Processing Units (IPUs) for deep learning applications in cosmology. It presents the benchmark between a Nvidia V100 GPU and a Graphcore MK1 (GC2) IPU on three cosmological use cases: a classical deep neural network and a Bayesian neural network (BNN) for galaxy shape estimation, and a generative network for galaxy images simulation. The results suggest that IPUs could be a potential avenue to address the increasing computation needs in cosmology.

\bigbreak
\tableofcontents
\bigbreak
\bigbreak
\section{Introduction}
\label{sec:introduction_appendix_ipu}
Upcoming imaging galaxy surveys, such as the Legacy Survey of Space and Time 
\citep[LSST,][]{2009arXiv0912.0201L} conducted at the future Vera C. Rubin Observatory, ESA’s Euclid satellite \citep{2010arXiv1001.0061R} or the Nancy-Grace-Roman Space Telescope \citep[Roman Space Telescope, former WFIRST,][]{2013arXiv1305.5422S} will produce an unprecedented amount of observational data. For instance, LSST will produce 20 Terabytes of data every night and around 60 Petabytes over the its 10 years of service\footnote{\url{https://www.lsst.org/about/dm}}. These amounts are mainly due to the quality and the nature of the data recorded: large sky images in different filters (or colors). 

In order to prepare and test future data analysis pipelines, it is necessary to simulate a quality and a quantity of data as close as what will be recorded.
The pipelines will need to process data with fast and accurate analysis techniques, and to provide reliable uncertainties. 
When looking for fast data processing techniques, a now common choice is to turn to deep learning algorithms, which allow for very fast inference on data once trained. 

As an example, a LSST pipeline will be dedicated to look for transients continuously during the night. It will have to send an alert to the community in within 60 seconds after image readout\footnote{\url{https://www.lsst.org/about/dm}}. The alert system is supposed to produce around 10 million alerts per night. A solution proposes to use deep learning to accelerate parts of the process such as the classification of events \citep{2020MNRAS.tmp.3384M}. 

In order to train deep artificial neural networks, specific hardware are necessary. Graphics Processing Units (GPU) are the most common technology used nowadays. It evolved a lot over the years to reach very high performance at the moment and surpasses Central Processing Units (CPU).
Google Tensor Processing Units (TPU), LightOn Optical Processing Units (OPU)\footnote{\url{https://www.lighton.ai/}}, or Graphcore Intelligence Processing Units (IPU)\footnote{\url{https://www.graphcore.ai/}}, are more recent examples of hardware developments aiming for faster computing. 

Graphcore's IPUs have already shown better results than GPUs at language and speech processing, computer vision, probabilistic modeling\footnote{\url{https://www.graphcore.ai/benchmarks}} and even in some common use cases of particle physics \citep{mohan2020studying}. 

\begin{figure*}
    \centering
    \makebox[\textwidth][c]{
    \includegraphics[trim={0.4cm 0.cm 0.3cm 0.cm}, clip, width=0.5\textwidth]{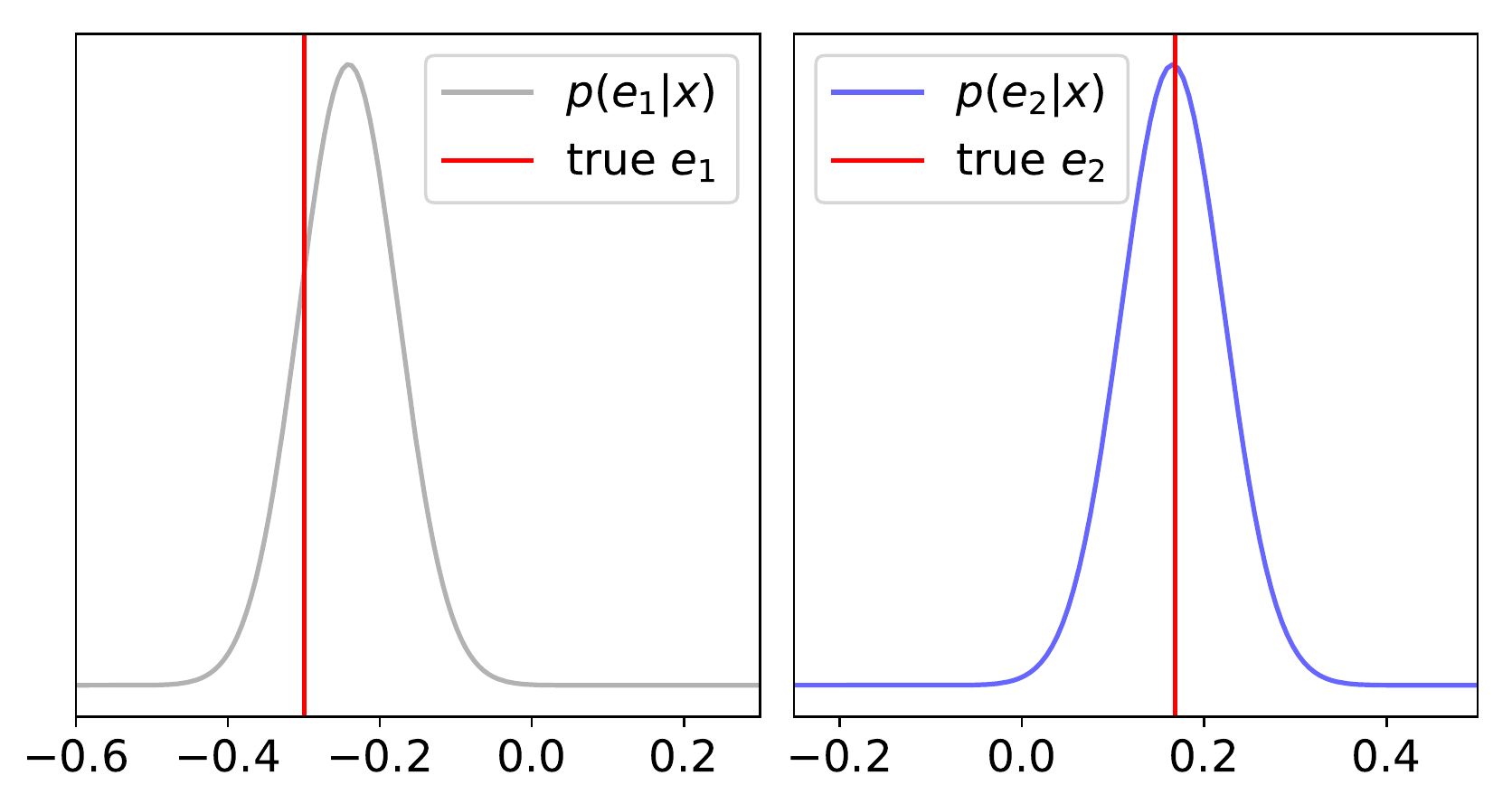}
    \includegraphics[trim={3.7cm 0.cm 0.3cm 0.cm}, clip, width=0.5\textwidth]{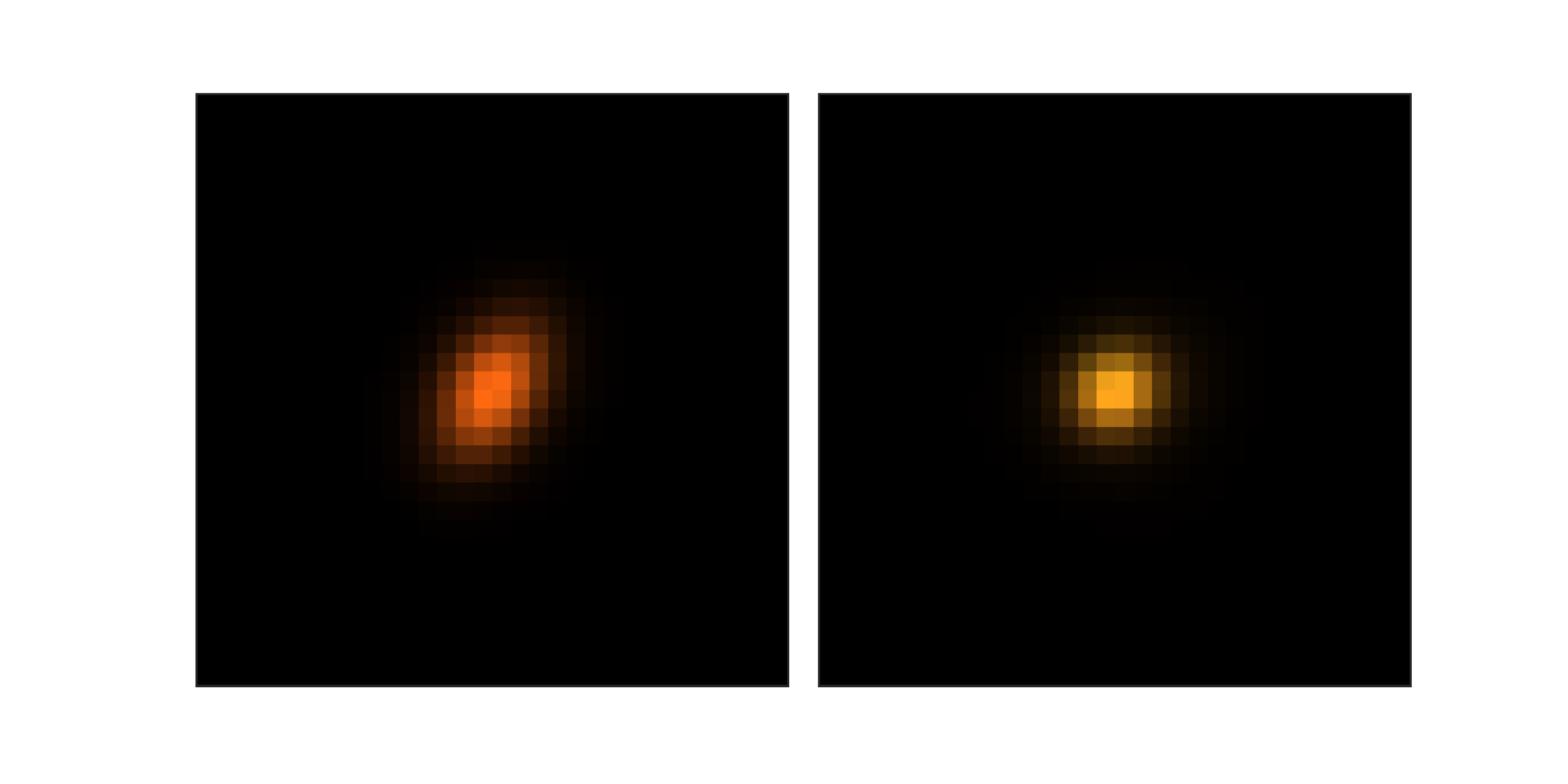}
    }
    \caption{On the left, the posterior distributions of a galaxy $e_1$ and $e_2$ shape parameters estimated with a deterministic deep neural network. On the right, two galaxy images generated via sampling the latent space of a trained VAE.}
    \label{fig:cosmo_examples}
\end{figure*}

This paper presents the benchmark between a Nvidia V100 GPU and a Graphcore MK1 (GC2) IPU on three cosmological use cases: a deterministic deep neural network and a Bayesian neural network (BNN) for galaxy shape estimation, and a generative network for galaxy images production. \Cref{fig:cosmo_examples} shows outputs of the previously mentioned use cases: the posterior distribution of galaxy ellipticity parameters, estimated from a deterministic network on the left, and isolated galaxy images generated from a neural network on the right. The code to reproduce this study can be found on GitHub \github \footnote{\url{https://github.com/BastienArcelin/IPU-GPU}}.

This paper represents a first investigation of the suitability and performance of IPUs in deep learning applications in cosmology. The paper is organised as follows: in \cref{sec:hardware_appendix_ipu}, the two tested hardware are described, \cref{sec:galaxy_shape_estimation_appendix_ipu} presents the results obtained to train a deterministic and a Bayesian neural network to learn galaxy shape parameters from isolated galaxy images, \cref{sec:galaxy_img_generation_appendix_ipu} compares the performance of the tested hardware in the case of inference, to generate isolated galaxy images, and finally, \cref{sec:conclusion_appendix_ipu} concludes and discusses the different results.

\section{Hardware description}
\label{sec:hardware_appendix_ipu}
In this paper I compare the performance of a single first generation MK1 (GC2) IPU\footnote{\label{note_ipu}\url{https://www.graphcore.ai/products/ipu}} to the performance of a single Nvidia V100 GPU. Key specifications of the IPU and the GPU\footnote{\label{note_gpu}\url{https://www.nvidia.com/en-gb/data-center/v100/}} are presented in \cref{tab:specs}. 

The IPU is a type of processor with a different architecture from the one of the GPU \citep{DBLP:journals/corr/abs-1912-03413}. It is specifically designed for machine learning applications as it offers true MIMD (Multiple Instruction, Multiple Data) parallelism. It is designed to adapt to irregular computation and sparse data access. Each processor is composed of $1216$ cores, called tiles. Each tile contains $256$ KiB of local memory and is complex enough to execute distinct programs. It can also support up to six threads, allowing for $7.296$ threads that can be executed in parallel on an IPU. On the contrary, GPU has a SIMD (Single Instruction, Multiple Data) architecture. The $5120$ GPU cores are grouped in clusters for which all of the cores will execute the same instruction. As threads are scheduled on clusters, they perform the same operation on independent data.

Two other differences have a direct impact for the user: the memory per processing unit and the single precision performance. The IPU has a much smaller memory than the GPU. This may become a bottleneck in case of large neural network, \ie with a lot of training parameters, or when processing heavy data, such as imaging survey data.

The other major difference is the single precision performance, IPUs being able to reach 31.1 TFLOPS (TeraFLOPS, Floating-point operation per second), more than twice the GPU value.

Finally, it is significant to notice that the IPU used here reaches the presented level of performance while consuming less than half electrical power consumed by the GPU. 

Note than more recent and more powerful versions of both Graphcore and Nvidia technologies have been released: the Colossus™MK2 GC200 at Graphcore, with increased in-processor and exchange memory, and the A100 GPU at Nvidia, with more memory and higher single precision performance.

Both GPUs and IPUs support TensorFlow \citep{2016arXiv160304467A} or PyTorch \citep{2019arXiv191201703P}, user-friendly frameworks allowing programmers without specific hardware knowledge to access high-performance computing. 

Examples presented in this work are implemented using TensorFlow in its version 2.1.0. Graphcore provided drivers and its Poplar Software Development Kit (SDK) which is used in its version 1.4.0. The GPU is accessed via the IN2P3 Computing Center (CC-IN2P3--Lyon/Villeurbanne - France), and runs with Nvidia CUDA version 10.1.105.

\begin{table}
    \makebox[\textwidth][c]{
    \begin{tabular}{c|c|c|c|c|c}
         & &  &  & Single precision & Max Power\\
         & Processing Unit & Cores & Memory & performance & Consumption\\
         \hline
         GPU & Nvidia Tesla V100 PCIe & 5120 & 32000Mb & 14 TFLOPS & 250 W\\
         IPU & Graphcore Colossus\texttrademark  GC2 & 1216 & 286Mb & 31.1 TFLOPS & 120 W\\
    \end{tabular}
    }
    \caption[Caption for LOF]{Comparison of IPU\cref{note_ipu} and GPU\cref{note_gpu} specification for machines used for this benchmark.}
    \label{tab:specs}
\end{table}

\section{Cosmological use cases}
\subsection{Training data}
Images generation uses the same hypothesis as what is described in section 2.1 of \cite{2021MNRAS.500..531A}. Images of isolated galaxy are simulated using \texttt{GalSim}\xspace\footnote{\url{https://github.com/GalSim-developers/GalSim}} \citep{2015A&C....10..121R} from parametric models fitted to real galaxies from the HST COSMOS catalogue, containing \num{81500} images. These fits were realised for the third Gravitational Lensing Accuracy Testing (GREAT3) Challenge \citep[see Appendix E.2]{2014ApJS..212....5M}. 

This benchmark is realised on simulations of LSST-like images composed of the six LSST filters ($ugrizy$, \cite{2009arXiv0912.0201L}) and convolved with a fixed Point-Spread-Function (PSF). For each filter, Poisson noise is added. Images size is arbitrarily fixed at $64 \times 64$ pixels and the pixel size is $0.2$ arcsecond. The code for image generation can be found here \githubimggen \footnote{\url{https://github.com/BastienArcelin/image_generation_GalSim}}.

Artificial neural networks presented in the next sections are trained by feeding them by batches. The batch size is not fixed, it is used as a parameter to compare the performance of the two tested hardware. The dimension of the input tensors is $(batch size, 6, 64, 64)$.

\subsection{Galaxy shape parameter estimation}
\label{sec:galaxy_shape_estimation_appendix_ipu}
With future galaxy surveys, various probes will be used to study dark energy. One of the most promising is cosmic shear \citep{2015RPPh...78h6901K}, \ie the measurement of the coherent distortion of background galaxies by foreground matter through weak gravitational lensing. To obtain a precise measurement of this probe, the shape and redshift (or distance) of the observed galaxies must be accurately measured. 
Several galaxy shape measurement methods based on model-fitting or moments measurement already exist, but none of them perform accurately on blended objects, \ie overlapping objects on an image. As galaxy surveys observe further in the sky, dealing with overlapping objects is becoming a major challenge. For example, \cite{Bosch2017} estimated that 58\% of detected objects will be blended in the HSC wide survey.

We developed a new technique based on deep neural networks and convolutional layers which permits to measure shape ellipticity parameters on isolated as well as on blended galaxies \citep{arcelininprep}. Our network takes as input the isolated galaxy or blended galaxies scene and outputs the posterior distribution of the target galaxy ellipticity parameters. First, we tested a deterministic neural network, and then a BNN. The advantage of BNNs is that they allow for an accurate characterisation of the epistemic uncertainty, which is necessary to do reliable prediction that can be used in scientific studies \citep{2020arXiv200601490C}.

As this work focuses on hardware performance, the network tested here is a simplified version of the one working on blended scenes and presented in \cite{arcelininprep}. It is trained to retrieve ellipticity parameters distribution from isolated galaxy images only. It is composed of one batchnormalization layer, eight stacked convolutional layers, one dense layer and the output is a multivariate normal distribution layer\footnote{\url{https://www.tensorflow.org/probability/}} from the TensorFlow Probability library.

\subsubsection{Deterministic neural network}
\label{sec:galaxy_shape_estimation_appendix_ipu_det}
The training of a deterministic network performing galaxy shape parameter estimation is the first test presented here. The network is defined as \textit{deterministic} since, once trained, trainable parameters have a fixed value. In other words, the output posterior distributions will not change if the network is fed twice with the same isolated galaxy image. This network is composed of $1.5$M parameters. 

During training, the Adam optimizer \citep{2014arXiv1412.6980K} is used and the loss is defined as the negative log probability of the output:
\begin{equation}
    \mathcal{L} = - log(p(x))
\end{equation}
Here $p$ is the posterior distribution, output of the multivariate normal distribution layer and $x$ is the target value.

The training procedure is the same for IPUs and GPUs and the chosen metric is the training time for 100 epochs. Ten epochs are considered as a warm-up phase. The test is realised for different batch sizes: from 2 to 14, 14 being the maximum batch size that can be handled by the IPU before memory is full (see \cref{sec:hardware_appendix_ipu}).

\begin{figure*}
    \centering
    \makebox[\textwidth][c]{
    \includegraphics[trim={0.cm 0.cm 0.3cm 0.cm}, clip, width=\textwidth]{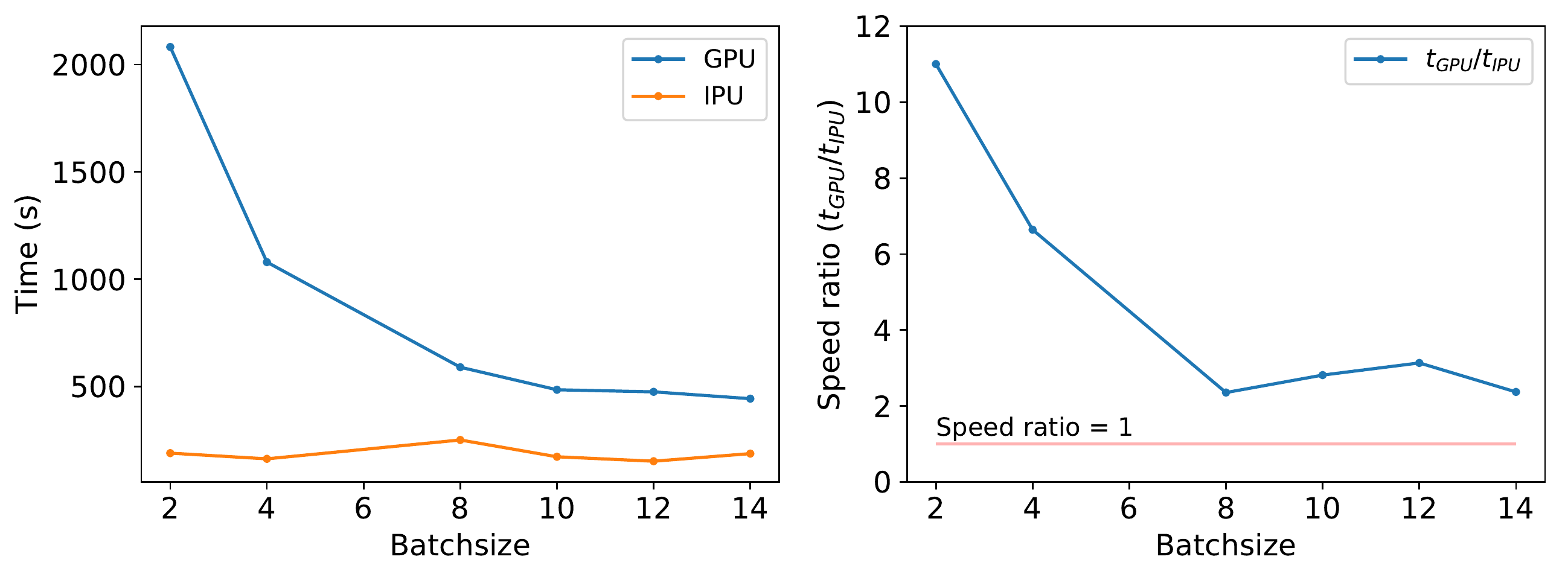}
    }
    \caption{(left)Training time for a deep deterministic neural network performing galaxy parameter estimation as a function of the batch size. (right) Speed ratio of GPU and IPU as a function of the batch size. The red horizontal line represents the value R = 1, \ie the network takes the same amount of time for training in both cases. The neural network is composed of $1.5$M trainable parameters.}
    \label{fig:training_time_deterministic}
\end{figure*}

The results are presented on \cref{fig:training_time_deterministic}. Using IPUs improves the training time compared to GPUs, especially for very small batch sizes. The network does not learn well for batch sizes lower than 4, but even for greater values, the network is trained at least twice as fast on the IPU compared to the GPU. This result confirms the already shown efficiency of IPUs on tasks close to computer vision\footnote{\url{https://www.graphcore.ai/mk2-benchmarks}}.

\begin{figure*}
    \centering
    \makebox[\textwidth][c]{
    \includegraphics[trim={0.cm 0.cm 0.3cm 0.cm}, clip, width=\textwidth]{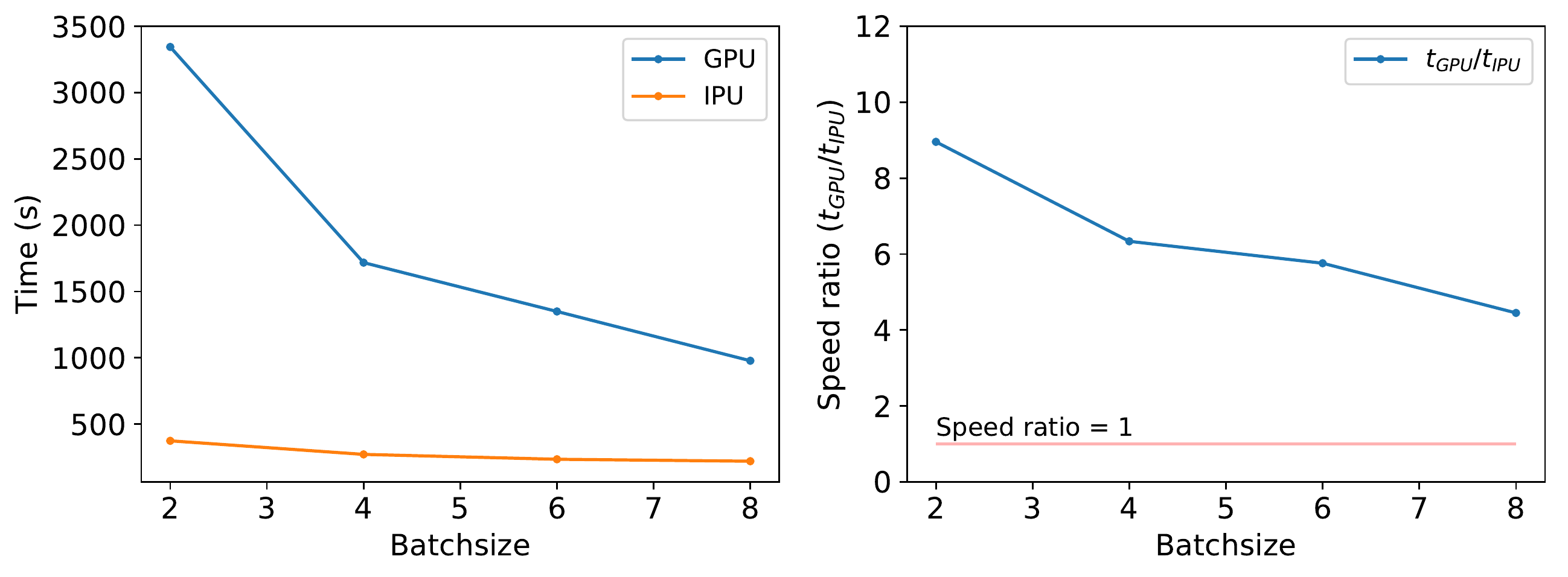}
    }
    \caption{(left)Training time for a deep BNN performing galaxy parameter estimation as a function of the batch size. (right) Speed ratio of GPU and IPU as a function of the batch size. The red horizontal line represents the value R = 1, \ie the network takes the same amount of time for training in both cases. The neural network is composed of $2.7$M trainable parameters.}
    \label{fig:training_time_bnn}
\end{figure*}

\subsubsection{Bayesian neural network}
\label{sec:galaxy_shape_estimation_appendix_ipu_BNN}
Then, the same test is run, but this time with a BNN. Instead of learning fixed values for each parameter, training a BNN consists in learning a posterior distribution over each parameter, knowing the data, through variational inference (\cite{2020arXiv200706823V} and \cite{2020arXiv200601490C}). Here the reparameterization trick is used to perform variational inference (\cite{2013arXiv1312.6114K}, \cite{2015arXiv150602557K} and \cite{2015arXiv150505424B}).

Since the training consists in learning distribution parameters, the number of trainable parameters increases to $2.7$M. Training BNNs also adds a term in the loss: the sum of all the Kullback-Leibler (KL) divergences between the parameter posterior distribution and the prior distribution. Here, this prior distribution is chosen to be the normal distribution, a common choice in literature. 
The loss is now defined as
\begin{equation}
    \mathcal{L} = -log(p(x)) + \sum_i{KL(p(w_i|x)||p(z))}
\end{equation}
Here $w_i$ is the $i-th$ parameter in the network and $p(z)$ is the prior distribution. 

Same as previously, this network is trained during 100 epochs on the same images sample, with a 10 epochs warm-up phase.

As expected, the memory size limitation restrains the training of the BNN to lower batch sizes than with the deterministic network. The BNN is trained with batch sizes varying from $2$ to $8$. IPUs once again outperform GPUs (see \cref{fig:training_time_bnn}) with a training speed at least four time superior.

To summarise, training artificial neural networks on IPUs is much faster than on GPUs. However, the memory size of IPUs becomes a limitation for large networks as it restrains their training to small batch sizes. This might become a problem in some cases, if the batch sizes are too small for the learning to converge. In order to go to higher batch sizes, it would be necessary to split the network over several IPUs.

\subsection{Galaxy image generation}
\label{sec:galaxy_img_generation_appendix_ipu}
Generating large amount of mock data is a major step to test and improve analysis pipelines. Until now, most galaxy simulations are based on simple analytic profiles such as Sérsic profiles \citep[see for example][]{2015A&C....10..121R}. This is not very fast and is known to increase the risk of introducing  model  biases \citep[see][]{2015MNRAS.450.2963M}.

Deep learning is a natural choice to turn to when looking for fast data generation methods. Several generative models based on neural networks already exist and present interesting performance (Variational Auto-Encoder \citep{2013arXiv1312.6114K}, Generative Adversarial Network \citep{2014arXiv1406.2661G} or Normalizing flow \citep{2014arXiv1410.8516D}, \citep{2015arXiv150505770J}). In our context, they are also effective tools to learn how to generate complex light profiles and decrease model bias \citep{2020arXiv200803833L}. Using generative neural networks to model galaxy is seriously considered in different cosmological applications (\cite{2020arXiv200803833L}, \cite{Regier2015ADG} or \cite{2021MNRAS.500..531A}).

\subsubsection{Generative model}
In this part, the images generation is done by sampling the latent space distribution of a variational autoencoder (VAE) trained on isolated galaxy images. When training the VAE, variational inference is used to learn a latent space distribution which encodes the galaxy parameters distribution. The prior for the latent space is taken as a normal distribution. However with training, the posterior distribution of the latent space is almost always not exactly normal (see Fig.3 of \cite{2021MNRAS.500..531A} for instance). It is then possible to use normalizing flows to map this distribution into a normal one thanks to a series of small invertible transformations. From there, it is possible to trivially sample the latent space distribution via the normalizing flows and use the decoder to generate galaxy images. I used Masked Autoencoder for Distribution Estimation \citep[MADE, ][]{2015arXiv150203509G} to build the normalizing flows network (similarly to \cite{2020arXiv200803833L}).
Here I compare image generation (or inference) speed using IPUs and GPUs. After a warm-up phase of 100 inferences, the measurement is realised on generating 1 to 5000 isolated galaxy images.

\begin{figure*}
    \centering
    \makebox[\textwidth][c]{
    \includegraphics[trim={0.cm 0.cm 0.3cm 0.cm}, clip, width=\textwidth]{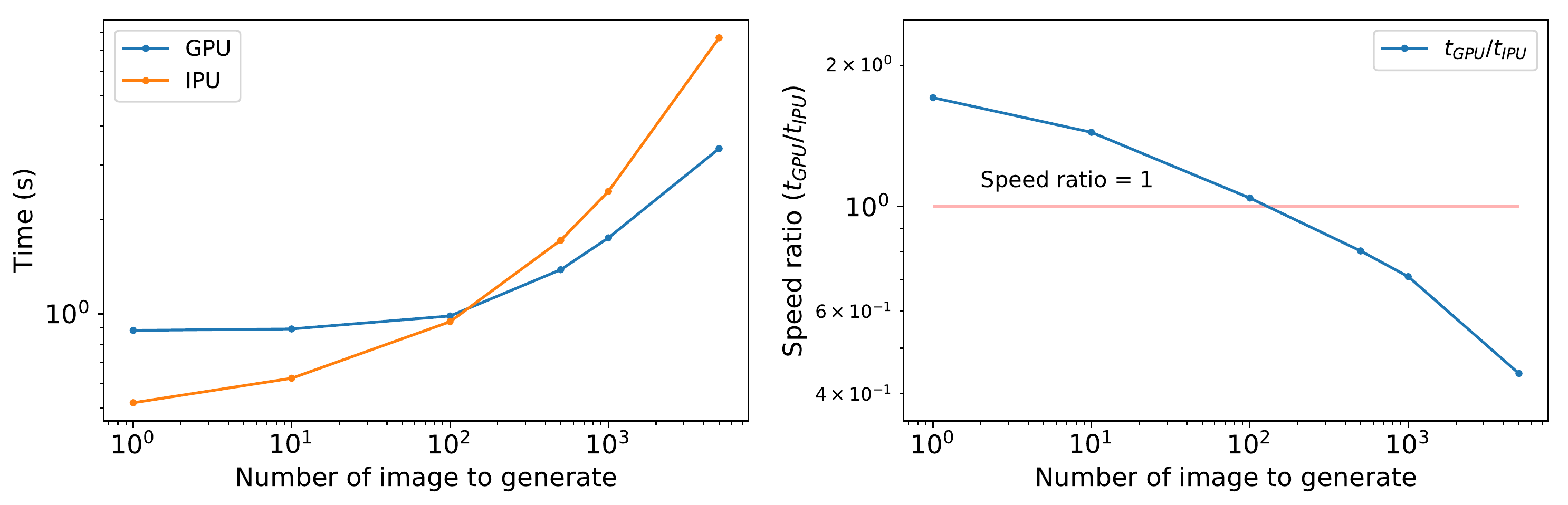}
    }
    \caption{(left) Inference time, sampling from a trained VAE latent space as a function of number of image to generate. (right) Ratio of inference time of GPU and IPU. The red horizontal line represents the value R = 1, \ie the network takes the same amount of time for inference in both cases. }
    \label{fig:inference_time_generation}
\end{figure*}

\subsubsection{Results}
\Cref{fig:inference_time_generation} shows that IPUs perform better than GPUs at generating small batches of images. On the contrary, GPUs outperform IPUs when the number of images to generate increases. The choice of hardware depends on the number of images to produce, which, in turn, can depend on the usage of the generated data. To train a network, generating small batches of data on the fly can be necessary and IPUs should be used.
On the contrary if the goal is to generate a large amount of data in order to test a data processing pipeline for example, generation on GPUs is more appropriate. As an example, the CosmoDC2 \citep{2019ApJS..245...26K} catalog released which covers $440$ $\deg^2$ of the sky area, contains around 2.26 billion galaxies. To generate this many galaxies with the tested generative model, less than 18 days of computing would be required on a single V100 GPU. Scaling this process on multiple GPUs would decrease this duration even more.

\section{Summary and discussion}
\label{sec:conclusion_appendix_ipu}
Parameter estimation and simulated images generation are typical examples of tasks that will be required to process and prepare for future imaging galaxy surveys. The increasing use of neural networks and particularly BNNs in cosmology demonstrates the community's interest in developing fast and accurate tools. These tools require particularly suited hardware. In this work I investigated IPUs and GPUs performance at training neural network and performing inference for cosmology applications.

In this study, we demonstrated that IPUs performed at least twice as fast as GPUs at training neural networks with small batch sizes. The restriction to small batch sizes due to the IPU memory size depends on the number of parameters of the network, leading to greater constraints for Bayesian neural networks. However, these constraints did not prevent the networks presented in this work to learn how to perform accurately.

We also presented performance in generating galaxy images from a trained VAE latent space, a typical inference example. Here, IPUs perform better at small batch sizes but are outperformed by GPUs at larger ones. The hardware choice depends on the task to accomplish: if the data is generated on the fly to train a neural network for example, using IPUs is probably more relevant. On the contrary, if the objective is to generate a large amount of data to test a data processing pipeline, it is more relevant to use GPUs. 

To conclude, this first test of IPUs for deep learning applications in cosmology suggests that IPUs perform better than GPUs at training neural networks but, regarding inference, the choice depends on the task to realise. It must be noted that the memory size of IPUs induces a limitation for training, in network size as well as in batch size, which might be a bottleneck for BNNs. This limitation might however be alleviated with the new version of Graphcore's IPUs. In any case, if these kind of processes are used to analyse cosmological data, scaling them on several IPUs or GPUs would probably be necessary, which might also solve the IPU memory size issue. 

\section{Acknowledgements}
I am grateful to Graphcore and to Microsoft for providing cloud access to their IPU through Azure IPU Preview as well as helpful technical support. I would also like to thank the IN2P3 Computing Center (CC-IN2P3--Lyon/Villeurbanne - France), funded by the Centre National de la Recherche Scientifique, for providing access to their GPU farm. This project was partially funded by the AstroDeep project\footnote{\url{https://astrodeep.net/}} from the Agence Nationale de Recherche (ANR).

\printbibliography

\end{document}